%% file: main.tex
\documentclass[conference]{IEEEtran}
\IEEEoverridecommandlockouts

\usepackage{orcidlink}

\usepackage{cite}
\usepackage{amsmath,amssymb,amsfonts}
\usepackage{algorithmic}
\usepackage{graphicx}
\usepackage{textcomp}
\usepackage{xcolor}
\usepackage{cleveref}
\usepackage{csquotes}
\usepackage{hyperref}
\usepackage{balance}
\usepackage{microtype}

\def\BibTeX{{\rm B\kern-.05em{\sc i\kern-.025em b}\kern-.08em
    T\kern-.1667em\lower.7ex\hbox{E}\kern-.125emX}}
\begin{document}

\input{custom-commands}

\title{ViMoTest: A Tool to Specify ViewModel-Based GUI Test Scenarios using Projectional Editing}

\author{\IEEEauthorblockN{Mario Fuksa \orcidlink{0000-0002-8210-094X}}
\IEEEauthorblockA{\textit{Institute of Software Engineering} \\
\textit{University of Stuttgart}\\
Stuttgart, Germany}
\and
\IEEEauthorblockN{Sandro Speth \orcidlink{0000-0002-9790-3702}}
\IEEEauthorblockA{\textit{Institute of Software Engineering} \\
\textit{University of Stuttgart}\\
Stuttgart, Germany}
\and
\IEEEauthorblockN{Steffen Becker \orcidlink{0000-0002-4532-1460}}
\IEEEauthorblockA{\textit{Institute of Software Engineering} \\
\textit{University of Stuttgart}\\
Stuttgart, Germany}
}

\maketitle

\begin{abstract}
Automated GUI testing is crucial in ensuring that presentation logic behaves as expected. 
However, existing tools often apply end-to-end approaches and face challenges such as high specification efforts, maintenance difficulties, and flaky tests while coupling to GUI framework specifics.
To address these challenges, we introduce the ViMoTest tool, which leverages Behavior-driven Development, the ViewModel architectural pattern, and projectional Domain-specific Languages (DSLs) to isolate and test presentation logic independently of GUI frameworks.
We demonstrate the tool with a small JavaFX-based task manager example and generate executable code.
\end{abstract}


\section{Introduction}

Graphical User Interfaces (GUIs) are essential in modern software systems by providing the surface for the primary interaction between users and the system.
As GUIs often contain complexity, developing high-quality software systems faces the challenge of ensuring that GUIs are reliable and behave correctly.
Automating testing brings the potential to validate GUI behavior, while current GUI testing approaches usually provide end-to-end testing.
Those end-to-end approaches do not only verify the presentation logic but also the GUI framework specifics and often need a complex setup to bring the System Under Test (SUT) into the desired state.

Existing GUI testing tools deal with challenges and limitations.
These challenges include high coupling between test suites and the SUT, development and maintenance efforts, time behavior, or fragility of tests~\cite{testing_junior_gui_functional_testing,nass_challengesRemain_guiTestAutomation,selenium_garcia_survey}.
For example, most approaches directly interact with GUI widgets, which needs robust GUI widget identification to avoid high efforts when the GUI implementation evolves.

\begin{figure}[b]
  \vspace{-3mm}
  \centering
  \includegraphics[width=0.82\linewidth]{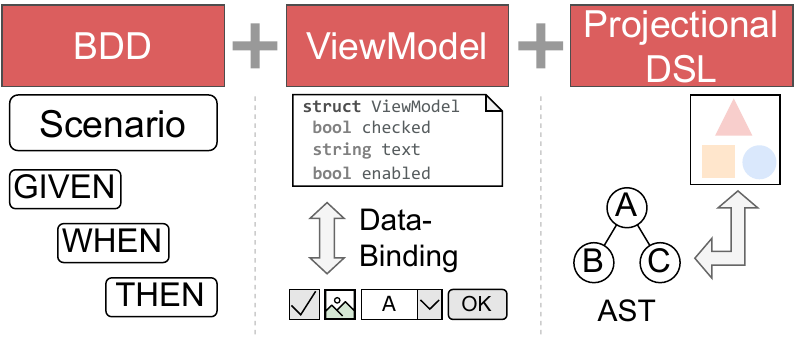}
  \vspace{-3mm}
  \caption{Three Combined Techniques of the ViMoTest Approach~\cite{fuksa_vimotest_docsym}.}
  \label{fig:vimotest_three_ingredients}
\end{figure}

To address those open challenges, we present a tool implementation of the ViMoTest approach~\cite{fuksa_vimotest_docsym}.
ViMoTest focuses on efficient test specification with explicit test oracles while decoupling GUI test scenarios from GUI frameworks.
The approach focuses on GUI application architectures with explicit testability design.
\Cref{fig:vimotest_three_ingredients} illustrates the three combined techniques: Behavior-driven Development (BDD), ViewModels, and projectional Domain-specific Languages (DSLs).
First, using the BDD technique, ViMoTest defines \textit{test scenarios}, which test developers write in a structured manner using the \textit{given-when-then} template.
Second, those test scenarios rely on \textit{ViewModels}\footnote{E.g., the ViewModel in the Model-View-ViewModel pattern acts as an intermediary between the View and the Model, exposing data and UI logic.}, providing a testable abstraction of a GUI's \textit{PresentationModel}~\cite{patterns_fowler_PresentationModel2004}.
ViMoTest specifically targets \textit{HumbleView/PresentationViewModel}~\cite{fuksa_mvvm} GUI architectures, where the ViewModel provides information in a format close to intended GUI widgets\footnote{E.g., a boolean for controlling the visibility feature of a widget.}.
This pattern supports focusing on the logical rather than the physical elements of the GUI, which reduces the coupling between tests and GUI implementations.
Third, ViMoTest uses projectional structured-code DSLs with the language workbench JetBrains MPS (\url{https://www.jetbrains.com/mps/}) to provide dedicated languages for describing logical widgets (e.g., buttons) with their supported features (e.g., visibility) and to specify test scenarios.
The projectional editing\footnote{Projectional editing skips traditional parsing by directly working with the abstract syntax tree, enabling notational freedom and language composition.} supports rendering logical widgets as intended by target GUI frameworks.
For example, when asserting a button is disabled, the test editor renders a grayed-out button in the scenario's then-part.
At the same time, MPS' editor supports features known from textual languages like auto-completion.

In an earlier research study, we validated the applicability of the ViMoTest concept by specifying test scenarios for five applications of the open-source domain using a preliminary prototype with limited generator support~\cite{fuksa_vimotest_applicability}.
As a successor artifact, we have implemented a more mature implementation, including an improved editor and code generation support in Java and C++.
Since we have not yet validated ViMoTest in real-world case studies, we demonstrate ViMoTest on a small JavaFX-based task manager project in this paper.
The task manager example demonstrates how the generated test scenarios from ViMoTest integrate into the project implementation.

\section{ViMoTest Concept}

\begin{figure}[t]
  \centering
  \includegraphics[width=0.92\linewidth]{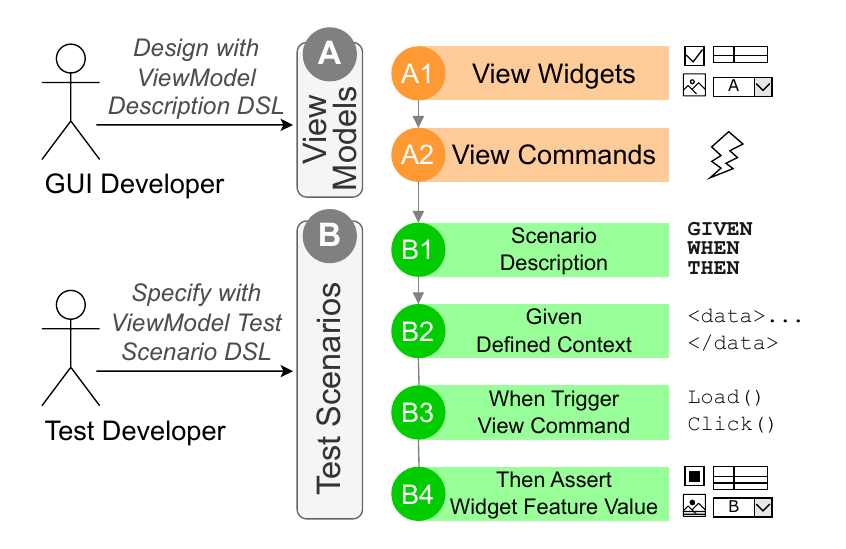}
  \vspace{-3mm}
  \caption{Envisioned Users and the ViMoTest Workflow~\cite{fuksa_vimotest_applicability}.}
  \label{fig:vimotest_workflow}
\end{figure}

ViMoTest addresses the goal of efficiently specifying robust test scenarios to test presentation logic.
\Cref{fig:vimotest_workflow} sketches the envisioned users: GUI developers and tester developers.
\roundIcon{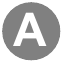} GUI developers design ViewModel descriptions containing logical widgets with supported features and view commands.
For example, using ViMoTest's \textit{ViewModel Description DSL}, the GUI developer describes a ViewModel with a checkbox widget, supporting a \textit{checked} feature and a \textit{enabled} feature, where a \textit{check command} supports controlling the checked-feature of the checkbox.
\roundIcon{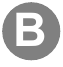} Based on those ViewModel descriptions, test developers specify test scenarios with the \textit{ViewModel Test Scenario DSL} to provide context, trigger ViewModel commands, and assert that the logical widgets have a certain state.
There might be further roles like domain experts, which can validate the test scenarios since they are specified in a language based on logical GUI widgets in a representation close to the target GUI.

\subsection{Task Manager ViewModel Example}

\begin{figure}[t]
  \centering
  \includegraphics[width=0.92\linewidth]{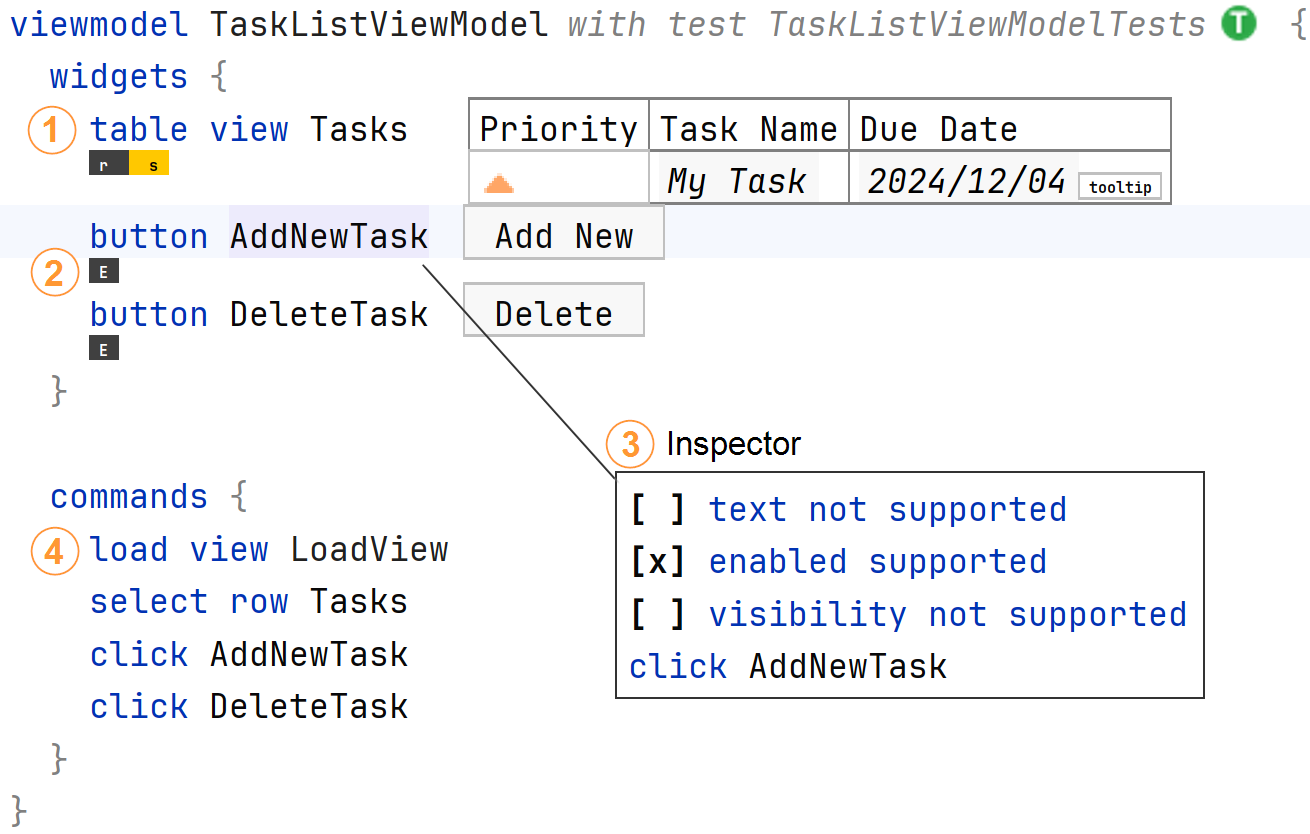}
  \vspace{-3mm}
  \caption{Task Manager ViewModel Example (ViewModel DSL).}
  \label{fig:example_taskview_viewmodel}
\end{figure}

We demonstrate the ViMoTest tool with a small task manager sample GUI, which consists of a table for listing tasks and two buttons for creating or deleting tasks.
We use ViMoTest to model and generate the API of the ViewModel and implement its presentation logic manually in a JavaFX project~\cite{zenodo_vimotest}.
\Cref{fig:example_taskview_viewmodel} shows the designed \texttt{TaskListViewModel} description in ViMoTest.
In the widgets section, the three logical GUI widgets are defined:
\numberIcon{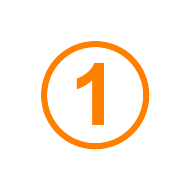}~we define a \textit{Tasks} table view widget, which consists of three columns \textit{Priority}, \textit{Task Name}, and \textit{Due Date}, where the first column defines image cells and the other two define label cells.
As depicted by the example, ViMoTest allows the insertion of example values for ViewModel descriptions.
The table also supports a selected row feature.
\numberIcon{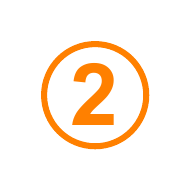}~The ViewModel additionally includes two button widgets \textit{AddNewTask} and \textit{DeleteTask}, which each support an \numberIcon{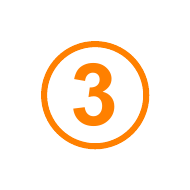}~enabled feature to let the presentation logic control their sensitivity.
\numberIcon{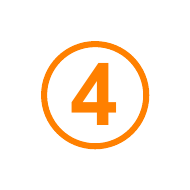}~For user interactions, we define four commands: a \textit{load view} command explicitly loads the ViewModel with context.
The \textit{select row} command controls the selection of the table widget.
Finally, two \textit{click commands} control the logic if one of the two buttons is clicked.

\subsection{Task Manager Test Scenarios}

\begin{figure}[t]
  \centering
  \includegraphics[width=0.90\linewidth]{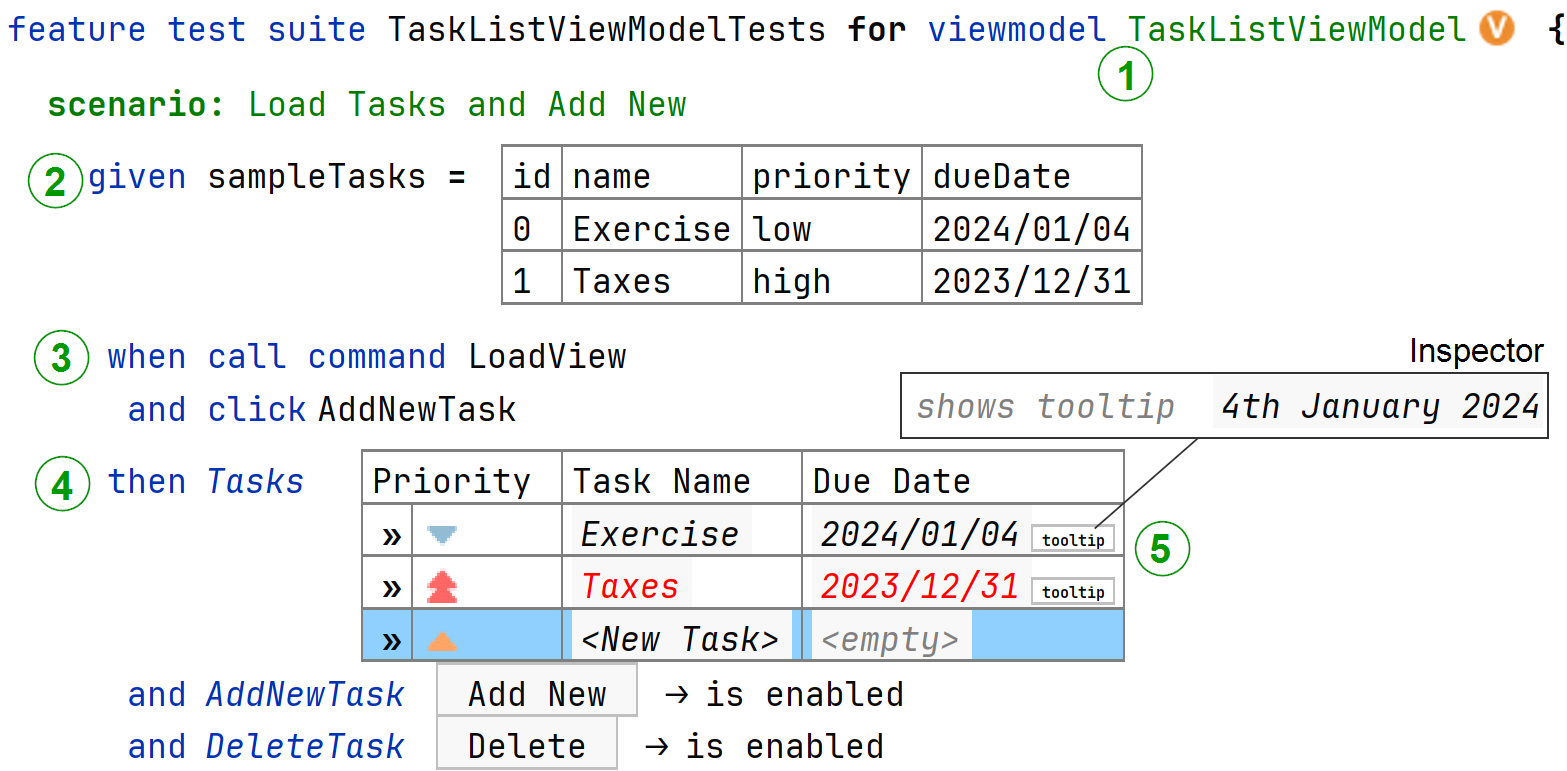}
  \vspace{-1mm}
  \caption{Task Manager Test Scenario Example (Test Scenario DSL).}
  \label{fig:example_taskview_test}
\end{figure}

Next, we specified test scenarios to test the three defined widgets of the SUT based on the \numberIcon{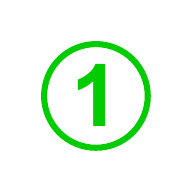}~\texttt{TaskListViewModel} definition.
\Cref{fig:example_taskview_test} shows a sample test scenario:

\begin{figure*}[t]
  \centering
  \includegraphics[width=0.9\textwidth]{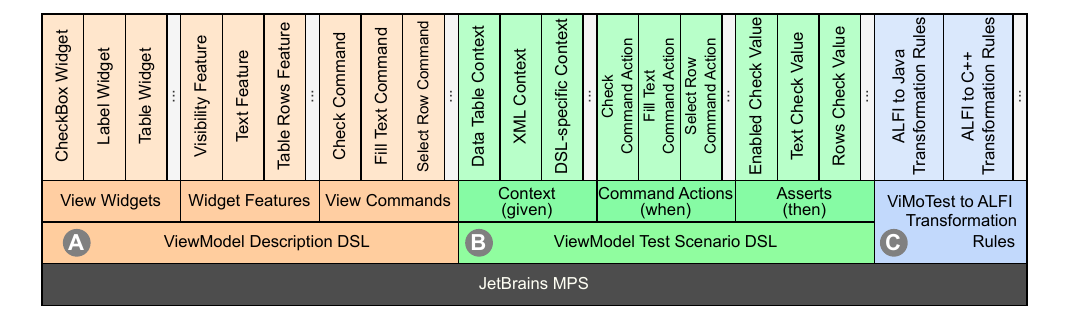}
  \vspace{-3mm}
  \caption{ViMoTest Tool Architecture.}
  \label{fig:vimotest_tool_architecture_boxes}
\end{figure*}

The scenario has the description \enquote{Load Tasks and Add New}, specified manually by the test developer.
\numberIcon{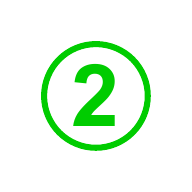}~The \textit{given} part defines a data table with four columns with the unique name \textit{sampleTasks}, where the first row specifies the header, and two further rows specify data values for two sample tasks.
We reused the idea of data tables from BDD to specify simple string-based tables.
\numberIcon{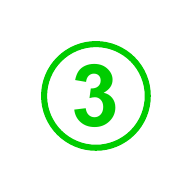}~The \textit{when} part contains a LoadView and an AddNewTask click command action.
The first command action explicitly triggers the ViewModel to fill widget features with the given context, while the second triggers the SUT to create a new task.
\numberIcon{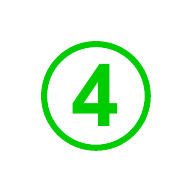}~Finally, the \textit{then} part contains the assertion of the \textit{Tasks} table and the two buttons.
\numberIcon{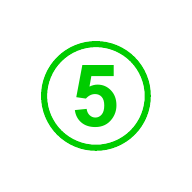}~We expect three rows with different priority image names, the task names \enquote{Exercise}, \enquote{Taxes}, and \enquote{New Task}, and the due dates according to the given data table, which also define a tooltip indicating due date in a different format, like \enquote{4th January 2024}.
The test verifies that the created row shall be selected and has an empty due date.
We also demonstrate the use of colors by expecting the second row to be marked red.

\subsection{Task Manager Code Generation}

Finally, we use the ViMoTest Java generator to transform the modeled ViewModel description and the test scenarios into executable Java code.
The generator produces an abstract Java class for the ViewModel implementation, which is then extended by a manually implemented class providing presentation logic.
The generator creates an executable JUnit test case for each test scenario, enabling automated verification of the implemented presentation logic.

\section{Implemented ViMoTest Tool Architecture}

This section describes the core elements of the ViMoTest tool architecture and provides insight into its implementation.

\subsection{ViMoTest Core Elements}

\Cref{fig:vimotest_tool_architecture_boxes} shows the architecture core elements.
First, we use the language workbench JetBrains MPS to develop the projectional structured-code ViMoTest DSL.
MPS does not rely on parsing but uses projections of the abstract syntax tree.
This supports notational freedom, which we leverage for rendering the ViewModel and expectations on logical GUI widgets representing the SUT.
Further, the projectional language workbench supports language composition, i.e., composing DSLs into host DSLs while the composed DSL is developed independently.
This option supports custom DSLs for context data definitions in the \textit{given} test scenario parts.

On top of MPS, we developed three main components: 

\aspectItalic{\roundIcon{Icon_A} ViewModel Description DSL}{
The first key component of the tool architecture is the \textit{ViewModel Description DSL}.
This DSL supports the design of the ViewModel's API, which is the shape of the SUT for generated test scenarios.
ViMoTest's ViewModel descriptions contain three aspects: \textit{View Widgets}, their supported \textit{Widget Features}, and \textit{View Commands}.

View widgets are logical elements independent of concrete GUI frameworks, representing placeholders for target GUI widget components.
Example element types are \textit{CheckBox Widget}, \textit{Label Widget}, or \textit{Table Widget}.

Each view widget defines features like \textit{Visibility Feature}, \textit{Text Feature}, or \textit{Table Rows Feature}.
Depending on the view widget type, features are inherently supported (e.g., a checkbox always has a Checked Feature) or optionally supported, i.e., the GUI developer decides if an optional feature is supported for each widget.

For supported user interactions, the ViewModel Description DSL covers \textit{View Commands}.
Examples of commands are \textit{Check Command}, \textit{Fill Text Command}, or \textit{Select Row Command}.
Most view commands are directly supported on View Widgets, like a \textit{Check Commands} supported on a \textit{CheckBox Widget}.
Depending on the specific type, those widget commands inherently define parameters, like a boolean for a \textit{Check Command}.
Besides widget commands, GUI developers can also define custom commands with custom parameter lists.
}

\aspectItalic{\roundIcon{Icon_B} ViewModel Test Scenario DSL}{
The second key component is the \textit{ViewModel Test Scenario DSL}.
This DSL enables test developers to specify {test scenarios} based on a ViewModel description.
Each test scenario contains the aspects \textit{Context (given)}, \textit{Command Action (when)}, and \textit{Asserts (then)}.

Like in BDD, \textit{Data Table Context} can define generic data in a tabular structure, which ViMoTest transforms into a multi-line string, JSON, or XML dependent on a separate configuration.
Additionally, the DSL supports low-level data formats like XML or multi-line strings.
However, the most domain-oriented context support can be reached with language engineering: an MPS developer can create custom sub-DSLs, which integrate into test scenarios by language composition.
Further, the DSL supports reference context definitions inside the same test suite to support reusability.

Next, the when part of a test scenario is defined by \textit{Command Actions}.
Command actions represent invocations of defined commands of the ViewModel description with concrete parameter values.
ViMoTest provides editor support for view commands, like a checkable box for a \textit{Check Command Action} or the string value for a \textit{Fill Text Command}.
For structural widgets as tables, \textit{Select Row Command Actions} support controlling the selection of the ViewModel's state.

The last part of test scenarios defines \textit{Asserts} containing check values to expect the ViewModel state in the SUT.
Examples of assertion types are \textit{Enabled Check Value}, \textit{Text Check Value}, or \textit{Rows Check Value}.
The test developer specifies whether to assert or ignore a supported feature on a view widget per test scenario.
By default, the editor renders tabular widgets that display all expected rows and cell values.
However, test developers can ignore entire columns or specific cell values, focusing only on relevant data in test scenarios.
}

\begin{figure}[t]
  \centering
  \includegraphics[trim=0mm 1mm 0mm 1mm, clip, width=0.99\linewidth]{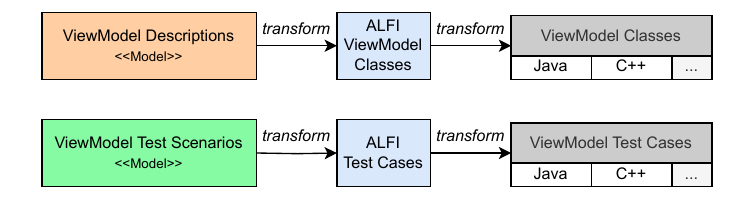}
  \vspace{-5mm}
  \caption{ViMoTest Transformation Chain.}
  \label{fig:vimotest_transformation_chain}
\end{figure}

\aspectItalic{\roundIcon{Icon_C} ViMoTest to ALFI Transformation Rules}{
The third key component of the ViMoTest tool architecture is about the \textit{ViMoTest to ALFI Transformation Rules}.
Since the goal of ViMoTest is to support multiple target programming languages, we introduced an intermediate level in the transformation chain.
For this intermediate level, we implemented ALFI (UML ALF as an Intermediate Language)~\cite{fuksa_alfi} to separate the ViMoTest-specific transformation rules from target language-specific transformation rules.
\Cref{fig:vimotest_transformation_chain} illustrates the ALFI-based transformation chain used in the ViMoTest tool architecture.
When writing this paper, ALFI supports transformation rules to Java and C++.
Therefore, ViMoTest also supports those two target languages.
We set up a continuous integration pipeline using GitHub Actions to compile the generated Java and C++ code of various modeled ViewModel descriptions and test scenarios to ensure the generated code is compilable.
}

\subsection{Solved Technical Challenges}

We solved technical challenges in developing ViMoTest.

\aspectItalic{Redundant Transformation Rules}{
ViMoTest supports multiple programming languages like Java or C++ (\Cref{fig:vimotest_transformation_chain}).
We implemented and successfully applied the intermediate MPS language \textit{ALFI}~\cite{fuksa_alfi} to avoid redundant transformation rules.
}

\aspectItalic{Meta-DSL for View Widgets}{
We observed in the earlier applicability study~\cite{fuksa_vimotest_applicability} when building a preliminary prototype that supporting widgets and features is a recurring effort in MPS since many MPS concepts must be created symmetrically to the architecture.
To reduce the effort and ensure consistency in creating new widgets and widget features, we implemented a small MPS-based \textit{Widget Meta-DSL}.
The \textit{Widget Meta-DSL} supports the definition of widgets, widget features, and commands and generates the necessary MPS meta-elements.
}

\aspectItalic{Name Bindings and Generation Configuration}{
ViMoTest test scenarios are specified against a ViewModel description shaping the SUT.
The generator produces target ViewModel classes, providing the API to which the generated test cases are linked.
However, when ViewModel classes are pre-implemented, and ViMoTest shall only generate test scenarios, the default names of the generated classes may not align with the existing ones.
We implemented \textit{name bindings} to face this issue so the GUI developer can annotate the ViewModel descriptions.
Those name bindings control different target names, such as type names, property names, getter operation names, or file names.

Further, we provide generation configuration settings to parameterize the transformation rules.
For example, the configuration controls if the ViewModel class contains commands, or a separate \textit{View Controller} class shall be generated.
Further, GUI developers can decide if the target ViewModel shall be generated as abstract classes and if the parameter object pattern shall be used for the generated command action statement.
}

\aspectItalic{Test Setup}{
ViMoTest defines context using data tables, XML, or custom DSL formats.
The transformation to actual test code needs to inject the generated context definition and connect it to statements in the given parts.
We designed \textit{Test Setup} classes, which the test developer manually implements to integrate with the application-specific SUT.
At the same time, ViMoTest uses a certain API of such class to pass the context definition data as string values.
We used two ways to pass those string values to the Test Setup.
We can transform file-based context definitions (e.g., an XML file) to an external file and pass the filename to the Test Setup.
Alternatively, we can transform the context to multi-line strings (e.g., for non-file XML or data tables) and pass them as in-line string values.
}

\subsection{Maturity Level and Requirements}

We have not yet validated the implementation of the ViMoTest tool in real-world case studies.
While it covers most of the initially collected requirements, it might lack the features when applied in real-world applications.
However, since we used an earlier preliminary prototype to experiment with, the current tool implementation is more mature and supports the generation of test scenarios in Java and C++.

We documented the high-level requirements for ViMoTest separately~\cite{zenodo_vimotest}.
The requirements result from the earlier ViMoTest applicability study~\cite{fuksa_vimotest_applicability}, and additional requirement elicitation meetings with industry experts.

\subsection{Novelty}

The ViMoTest concept combines BDD, ViewModels, and projectional structured-code editing in a novel way.
Other than established end-to-end GUI tools like Selenium or Playwright, ViMoTest uses ViewModels as the abstraction level of the DSL and addresses the technical domain of logical GUI widgets as first-class citizens without coupling to concrete GUI frameworks.
At the same time, projectional editing supports rendering the logical GUI widgets in test scenario assertions in a form the target GUI intends to render.
While ViMoTest uses the BDD technique to structure test scenarios with given-when-then verbs, it combines BDD directly with components of the language, such that no glue code has to be written as in Cucumber or keyword-driven testing approaches.

In contrast to our earlier applicability study~\cite{fuksa_vimotest_applicability}, which relied on a preliminary prototype and primarily emphasized the modeling aspect of ViMoTest, this tool paper presents the advanced, re-implemented architecture of ViMoTest.
It demonstrates a flexible test generation approach enabled by ALFI~\cite{fuksa_alfi}.
This new contribution highlights the generation of executable test cases tailored to realistic target GUIs, advancing the practical applicability of the ViMoTest framework.

There is research related to ViMoTest that we further differentiate from.
Patkar et al. propose a low code approach for BDD processes, where stakeholders can visually compose behavior tests~\cite{bdd_patkar_lowcode}.
Similarly, Zameni et al. propose an approach for automatic test generation and execution of BDD scenarios~\cite{bdd_zameni_testcase_gen}.
However, neither approach focuses on GUI testing.
Silva proposes an approach similar to the ViMoTest approach where a DSL describes BDD interaction scenarios for GUI testing~\cite{silva_dsl_interaction_scenarios_web_guis}.
Similarly, Bünder and Kuchen present an approach combining BDD, DSLs, and GUI testing~\cite{bundner_mdsd_bdd_guitesting}.
In contrast to those approaches, ViMoTest explicitly uses projectional DSLs supporting notational freedom and uses the ViewModel architectural pattern for code generation.

\section{Validation Plan}

We have not yet validated the ViMoTest tool as a research prototype in real-world case studies.
We will validate the tool in the near future by applying it in industrial case studies and open-source applications, where we will utilize a Goal-Question-Metrics plan.
These case studies will cover different situations.
First, practitioners will develop new GUIs and design ViewModels using the ViMoTest testing approach.
Second, we will validate the tool on existing applications where practitioners develop the ViewModels independently of ViMoTest.
Hence, the test scenario generation has to be aligned to the existing ViewModels, e.g., by name-bindings.

\section{Limitations and Extensions}

We identify the architectural requirement of using ViewModels in the GUI implementation as a tool limitation.
Although Model-View-ViewModel is an established GUI architectural pattern~\cite{fuksa_mvvm}, this prerequisite limits the ViMoTest approach to ViewModel-based GUI implementations.
Testing ViewModels also limits the scope to presentation logic, excluding GUI framework-specific rendering logic.

Further, ViMoTest provides DSLs that support a range of predefined GUI widgets and features.
While we identified in our earlier study~\cite{fuksa_vimotest_applicability} that most GUI implementations share common logical components, there might be a lack of support for special GUI widgets or features.
If required, language engineers can extend the JetBrains MPS-based tool, which requires specific know-how of the underlying platform.

While ViMoTest provides standard context formats like Data Tables, XML, JSON, or plain text in the \textit{given} parts, domain-specific context formats can be extended only by building JetBrains MPS sub-languages.
This also requires specific know-how of the underlying platform.

Also, when using ViMoTest for existing ViewModels, developers need to spend effort in configuring name bindings to make generated code compatible with generated tests.

\section{Tool Availability}

ViMoTest is an open-source tool, licensed under Apache 2.0, developed using JetBrains MPS~\cite{zenodo_vimotest}.

\balance
\section{Conclusion}

This paper presented the ViMoTest tool, designed to address challenges in testing the presentation logic of complex GUI applications like high specification efforts.
The concept of ViMoTest is about the roles of GUI developers and test developers, which can use dedicated DSLs to specify ViewModel descriptions and ViewModel Test Scenarios.
We designed the DSLs with a projectional structured-code language workbench in mind to represent the abstraction of logical GUI widgets in a visual way.
The testing DSL integrates the Behavior-driven Development (BDD) technique to structure test scenarios using the given-when-then terminology.
The ViMoTest tool architecture integrates concrete DSL implementations based on the language workbench JetBrains MPS and transformation rules to generate executable code in multiple target programming languages like Java or C++.

Future researchers can use the ViMoTest tool to study the approach of combining BDD, ViewModel abstractions, and projectional languages.
Practitioners can use the ViMoTest implementation to specify test scenarios with the provided DSL artifacts and generate code for executable test suites.

In the near future, we will validate ViMoTest by applying the tool to case studies of real-world applications in industry and the open-source domain.
We will share insights from those case studies in future research papers and a more in-depth conceptual discussion.

\bibliographystyle{IEEEtran}
\bibliography{IEEEabrv,main.bib}

\end{document}

%% file: custom-commands.tex
\newcommand{\aspect}[2]{
\noindent
\textbf{#1}: #2
}
\newcommand{\aspectItalic}[2]{
\noindent
\textit{#1}: #2
}

\newcommand{\aspectWithIndent}[2]{
\noindent
\hfill\begin{minipage}{\dimexpr\linewidth-0.4cm}
\textbf{#1}: #2
\end{minipage}
}

\newcommand{\aspectBox}[3]{
\noindent
\fbox{\begin{minipage}{0.96\linewidth}
\textbf{#1} #2: \textit{#3}
\end{minipage}}
}

\newcommand{\simpleBox}[1]{
\noindent
\fbox{\begin{minipage}{0.96\linewidth}
\textit{#1}
\end{minipage}}
}

\newcommand{\problem}[3]{
\noindent
\textbf{#1} - \textit{#2}: #3
}
\newcommand{\hypothesis}[1]{
\vspace{1mm}
\noindent
\textit{Hypothesis}: \textit{#1}
\vspace{1mm}
}

\newcommand{\researchquestion}[2]{
\noindent
\fbox{\begin{minipage}{0.96\linewidth}
\textbf{#1:} \textit{#2}
\end{minipage}}
\vspace{2mm}
}

\newcommand{\contribution}[3]{
\noindent
\fbox{\begin{minipage}{0.96\linewidth}
\textbf{#1} #2: \textit{#3}
\end{minipage}}
}

\newcommand{\rectIcon}[1]{\includegraphics[height=1.5ex,trim=1mm 3mm 1mm 1mm]{graphics/util/#1.pdf}}
\newcommand{\roundIcon}[1]{\includegraphics[height=1.5ex,trim=1mm 3mm 1mm 1mm]{graphics/util/#1.pdf}}

\newcommand{\numberIcon}[1]{\includegraphics[height=1.5ex,trim=8mm 8mm 8mm 8mm]{graphics/util/#1.pdf}}